\documentclass[aps,prc,superscriptaddress,twoside,twocolumn,nofootinbib,10pt,%
showpacs,floatfix]{revtex4-1}

\usepackage{amsmath,amssymb}
\usepackage{graphicx,bm}
\usepackage{slashed}
\usepackage{epstopdf}
\usepackage{ulem} 
\usepackage[usenames]{color}
\usepackage{float}

\renewcommand\sout{\bgroup \color{red} \ULdepth=-.5ex \ULset}

\begin{document}

\title{Baryon-baryon interactions at short distances \\
--  constituent quark model meets lattice QCD  -- }

\author{Aaron Park}
\email{aaron.park@yonsei.ac.kr}
\affiliation{Department of Physics and Institute of Physics and Applied Physics, Yonsei
University, Seoul 03722, Korea}
\author{Su Houng Lee}
\email{suhoung@yonsei.ac.kr}
\affiliation{Department of Physics and Institute of Physics and Applied Physics, Yonsei
University, Seoul 03722, Korea}
\author{Takashi Inoue}
\email{inoue.takashi@nihon-u.ac.jp}
\affiliation{Nihon University, College of Bioresource Sciences, Kanagawa 252-0880, Japan}
\affiliation{Quantum Hadron Physics Laboratory, RIKEN Nishina Center, Wako 351-0198, Japan}
\author{Tetsuo Hatsuda}
\email{thatsuda@riken.jp}
\affiliation{Interdisciplinary Theoretical and Mathematical Sciences Program (iTHEMS), RIKEN, Wako 351-0198, Japan}
\affiliation{Quantum Hadron Physics Laboratory, RIKEN Nishina Center, Wako 351-0198, Japan}

\begin{abstract}
The interaction energies between two baryons at short distance in different flavor channels  are calculated  from the  constituent quark model (CQM)
 and are compared with the recent lattice QCD (LQCD) results for baryon-baryon potentials at short distance.
  We consider  the six-quark system with two strange quarks and focus on the  quantum numbers,  (Flavor,Spin)=(1,0),(8,1),(10,1),($\overline{10}$,1) and (27,0).
The interaction energy  is defined by subtracting out isolated baryon masses and  relative kinetic energy of two baryons from the total energy of
 a compact six-quark state. We introduce  interaction energy ratio between different flavors as a  useful measure
  to test the prediction of CQM.  We find that the ratios in CQM show
 good agreement with those in LQCD, which indicates that
 the short range part of the baryon-baryon interaction can be  understood qualitatively in terms of the Pauli principle and spin-dependent color interaction
  among constituent quarks.
\end{abstract}

\maketitle


\section{Introduction}

Understanding the baryon-baryon (BB) interactions at short distances is not only important to
 search for possible dibaryon states but also to study the central core of neutron stars.
  The recent analyses of the flavor dependence of the BB interactions from first principle lattice QCD (LQCD) simulations near the physical quark masses
   indicate that their behavior at short distances (reviewed in \cite{Hatsuda:2018nes})
    are qualitatively consistent with the idea of the constituent quark model (CQM) where  the Pauli exclusion among quarks combined with the
  single gluon exchange play important roles (reviewed in \cite{Oka:2000wj}).
However, the quantitative comparison between the LQCD results and CQM results has not been
 conducted so far. In this paper, we carry out such a comparison by focusing our attention
   on the ``ratios"  instead of the absolute magnitudes of the BB potentials at short distances:
    We expect that
    the interpolating operator dependence in LQCD   as well as quark wave function dependence
     in CQM are cancelled separately in each case, so that the comparison can be made with less ambiguities
   in the ratio.
 As we will see, the color-spin-flavor structure together with the spin-dependent color interaction between quark pairs   in the six-quark states  provides insights into the origin of the repulsion or attraction in different flavor configurations.

This paper is organized as follows.  In Sec.II, we recapitulate the essential features of the CQM which was
    employed
    to study multi-quark systems by two of the present authors  \cite{Park:2016cmg,Park:2018ukx}.
     In Se.III, we introduce appropriate coordinate systems to describe six-quark system containing  two strange quarks.
 In Sec.IV, we evaluate the color-spin interactions in CQM for six-quark systems  with flavor SU(3) symmetric case
       and compare the results with the corresponding  LQCD data for heavy quark masses.
      In Sec. V,   we evaluate the interaction energies between two baryons at short distance  in CQM with flavor SU(3) non-symmetric case
       and compare the results with the corresponding  LQCD data for nearly physical quark masses.
   Sec. VI is devoted to summary and concluding remarks.

\section{Compact six quarks in CQM}

Let us consider a compact six-quark configuration from the CQM point of view.  We will assume that the spatial wave function of all the six quarks are in the lowest energy s-wave state.
Because the quarks have color, flavor and spin, only specific total quantum numbers are allowed for the  six quark configurations made of  two octet baryons.
 Under flavor SU(3) symmetry, the allowed states are
$ (F,S)=(1,0),(27,0),(10,1),(\overline{10},1),(8,1)$, where $F$ and $S$ are the irreducible flavor representation and the total spin, respectively.
 In the following, we will focus on the particular isospin states
with two strange quarks,
  \begin{align}
 F_{\ell} &\equiv (F,I,S)    \nonumber \\
 &=(1,0,0),(27,0,0), (10,1,1), (\overline{10},1,1),(8,0,1),
 \label{eq:FS}
\end{align}
 which are relevant  for making a direct comparison
 between the CQM results  and  the recent LQCD  results \cite{Inoue:2016qxt}.
 Hereafter,  we label the states in  Eq.(\ref{eq:FS}) by $\ell = 1, 27, 10, \overline{10}, 8$.

The interaction of two baryons at short distances in CQM \cite{Oka:2000wj}
 are governed by the quark dynamics with the following Hamiltonian,
\begin{eqnarray}
H=\sum_{i=1}^{N}(m_{i}+\frac{\textbf{p}^2_i}{2m_i})
 -\frac{3}{16}\sum_{i<j}^{N}(V^{C}_{ij}+V^{CS}_{ij}),\label{Hamiltonian}
\end{eqnarray}
where $N$ is the total number of constituent quarks and  $m_i$'s are the constituent quark masses.
The spin-independent (spin-dependent)  color interaction denoted as
$V^{C}_{ij}$ ($V^{CS}_{ij}$) is given by  \cite{Bhaduri:1981pn,Park:2016mez}.
\begin{eqnarray}
V^{C}_{ij}=  -  \lambda^c_i\lambda^c_j \left( - \frac{\kappa}{r_{ij}}+\frac{r_{ij}}{a_0}-D \right) ,
\end{eqnarray}
\begin{eqnarray}
 V^{CS}_{ij}
=\frac{{\kappa}^{\prime}}{m_im_j r_{0ij}^2 } \frac{1}{r_{ij}}e^{-(r_{ij}/r_{0ij})^2} \lambda^c_i\lambda^c_j \ {\sigma}_i\cdot{\sigma}_j ,
\end{eqnarray}
where
$\lambda^c_i$ are the Gell-Mann matrices of the $i$'th quark for the color SU(3).
Here, $r_{ij}$ is the distance between quarks, while $r_{0ij}$ is chosen to depend on
the constituent quark masses as
\begin{eqnarray}
r_{0ij}=  (\alpha+\beta \mu_{ij})^{-1}. \label{mass-cs}
\end{eqnarray}
with $\mu_{ij}=  {m_im_j}/({m_i+m_j})$ being the
reduced mass.\footnote{The  quark mass dependence in $r_{0ij}$ is introduced to fit the hadrons masses
 not only with light flavors but also with charm and bottom in the s-wave \cite{Park:2016mez}.}
Throughout this paper, we assume isospin symmetry, $m_u=m_d$.
 In Table \ref{parameters}, we show the values of the parameters fitted to low-lying baryon masses  with $S$=1/2 and 3/2 including those with charm and bottom
quarks \cite{Park:2016mez}.   Table \ref{parameters-2} shows  the calculated masses of
 light baryons relevant to this work.

\begin{center}
\begin{table}[t]
  \begin{tabular}{|c|c|c|c|c|c|c|c|c|c|}
     \hline
    $\kappa$  & $\kappa '$ & $a_0$  & $D$ & $\alpha$ &$\beta$ & $m_{u,d}$  & $m_s$ & $m_c$ & $m_b$   \\
    \hline \hline
    0.59  & 0.5 & 5.386 & 0.96  &  2.1    & 0.552 & 0.343  & 0.632 & 1.93 & 5.3 \\
          &     & $ \mbox{GeV}^{-2}$ & $\mbox{GeV}$ & $\mbox{fm}^{-1}$  & & $\mbox{GeV}$   & $\mbox{GeV}$ & $\mbox{GeV}$ & $\mbox{GeV}$\\
    \hline
  \end{tabular}
   \caption{The  parameters of CQM fitted to light and heavy baryon masses \cite{Park:2016mez}. }
  \label{parameters}
\end{table}
\end{center}

\begin{center}
\begin{table}[t]
   \begin{tabular}{|c|c|c|c|c|c|c|c|}
  \hline
  $(I,S)$      & ($\frac{1}{2}$,$\frac{1}{2}$) &($\frac{1}{2}$,$\frac{3}{2}$)& (0,$\frac{1}{2}$) & (1,$\frac{1}{2}$) &(1,$\frac{3}{2}$) &($\frac{1}{2}$,$\frac{1}{2}$) &($\frac{1}{2}$,$\frac{3}{2}$)    \\
           & N, P     & $\Delta$    & $\Lambda$   & $\Sigma$     & $\Sigma^*$   & $\Xi$    &  $\Xi^*$     \\
  \hline \hline
  $M_B$  & 0.977 & 1.23 & 1.12 & 1.2 & 1.38 & 1.324 & 1.52      \\
  \hline
  Expt.       & 0.938   & 1.232  & 1.115  & 1.189 & 1.382 & 1.315 & 1.532  \\
  \hline
  \end{tabular}
  \caption{Masses of light baryons relevant to the present work in the unit of GeV. }
  \label{parameters-2}
\end{table}
\end{center}

The static interaction energy $V_{\rm CQM}$ between two baryons located on top of each other
can be estimated by looking at the interaction in terms of the six-quark configuration relative to the   two-baryon threshold:
\begin{eqnarray}
   V_{\rm CQM} &=&  \langle H \rangle_{6q}  - E_{BB'} , \label{binding-V}  \\
  E_{BB'} &=&  M_{B} +M_{B'} + K_{{\rm rel}, BB'} . \label{binding}
\end{eqnarray}
Here
$\langle H \rangle_{6q}$ is the expectation value  of the Hamiltonian with $N=6$  with respect to the six-quark in the s-wave,
$M_{B}$ and $M_{B'}$  are the single baryon energies obtained by the Hamiltonian with $N=3$,
 and  $K_{{\rm rel}, BB'}$ is the relative kinetic energy between two baryons.
This formula will be used to estimate the interaction energies for both  the flavor SU(3) symmetric  and
non-symmetric cases.

The total energy of the six-quark system in CQM can also be decomposed as
\begin{eqnarray}
  \langle H \rangle_{6q} = \sum_{i=1}^6 m_i +  K + E_C + E_{CS} ,
  \label{eq:H6q}
\end{eqnarray}
 where $K$ stands for the total kinetic energy, $E_C$ is obtained from $V_C$  and $E_{CS}$ is obtained from $V_{CS}$.

\begin{center}
\begin{table}
  \begin{tabular}{|c|c|c|c|}
      \hline
      & $i,j$=1-4 & $i$=1-4,\ $j$=5,6 & $i,j=5,6$  \\
    \hline \hline
    $F_1$  & $-\frac{7}{6}$ & $-\frac{11}{12}$ & $-\frac{5}{3}$ \\
    \hline
    $F_{27}$  & $-\frac{5}{6}$ & $-\frac{17}{12}$ & $\frac{1}{3}$ \\
    \hline
    $F_{10}$ & $-\frac{35}{27}$ & $-\frac{13}{18}$ & $-\frac{22}{9}$ \\
    \hline
     $F_{\overline{10}}$  & $-\frac{31}{27}$ & $-\frac{17}{18}$ & $-\frac{14}{9}$ \\
    \hline
    $F_8$  & $-\frac{4}{3}$ & $-\frac{2}{3}$ & $-\frac{8}{3}$ \\
    \hline
  \end{tabular}
  \caption{Expectation value of the spin-independent color factor $\langle \lambda^c_i \lambda^c_j \rangle$
   of a quark pair in each flavor state.   $i=1-4$ label the light quarks and $i=5,6$  the strange quarks. }
  \label{color-color}
\end{table}
\end{center}

The average matrix elements for the quark pairs contributing to  the spin-independent color interaction $V_C$ are given in Table \ref{color-color}.   For flavor SU(3) symmetric case with $m_s=m_u$, this interaction  will not contribute in appreciable strength
 in Eq.(\ref{binding-V}) as long as  the spatial size of a single baryon and that of the 6-quark system similar.
 This is so because   the color factor in $V_C$ taken with respect to the color-singlet state
 is proportional to $N$ as
 \begin{eqnarray}
\sum_{i<j} \lambda_i^c \lambda_j^c  & = & -\frac{8}{3}N .
\end{eqnarray}
This is also true for   $f$-type or $d$-type  three-quark interactions  in the color singlet state  \cite{Park:2018ukx}:
$\sum_{i \neq j\neq k} f^{abc} \lambda^a_i \lambda^b_j \lambda^c_k =  0$ and
$\sum_{i \neq j\neq k} d^{abc} \lambda^a_i \lambda^b_j \lambda^c_k  =  \frac{160}{9} N $ where $f$ and $d$ are the antisymmetric and symmetric structure constants for color SU(3),  respectively.

On the other hand,  the sum of the color-spin factor in the color-spin interaction $V_{CS}$  depends non-linearly
on $N$ as will be shown in Sec.\ref{sec:CSI}, so that  it induces non-vanishing contribution to Eq.(\ref{binding-V}).

\section{Coordinate system for six quarks}

In the following, we consider   $V_{\rm CQM}$ for the six-quark system
 with two strange quarks as mentioned.
 To calculate $\langle H \rangle_{6q}$  in Eq.~(\ref{eq:H6q})  for different flavor states, we first construct
 the orbital-color-flavor-spin wave function for six-quark systems \cite{Park:2015nha}.  Since we place two strange quarks on the fifth and sixth positions,
 the total wave function of the compact six quarks  should obey the \{1234\}\{56\} symmetry
where the curly bracket means antisymmetric combination.
 We choose the spatial part of the wave function $|R \rangle$ satisfying [1234][56] with the square bracket
 being  symmetric combination,
 \begin{align}
|R \rangle = \frac{1}{\sqrt{\cal N}} e^{-a_1 (\textbf{x}_1^2+\textbf{x}_2^2+\textbf{x}_3^2)-a_4 \textbf{x}_4^2 - a_5 \textbf{x}_5^2}.
\label{eq:a145}
\end{align}
Here  ${\cal N}$ is a normalization factor and $\textbf{x}_{1, \cdots, 5}$ are the Jacobi coordinates,
\begin{eqnarray}
\textbf{x}_1 &=&\frac{1}{\sqrt{2}}(\textbf{r}_1-\textbf{r}_2), ~~~~~\textbf{x}_2=\frac{1}{\sqrt{6}}(\textbf{r}_1+\textbf{r}_2-2\textbf{r}_3) \nonumber \\
\textbf{x}_3 &=&\frac{1}{\sqrt{12}}(\textbf{r}_1+\textbf{r}_2+\textbf{r}_3-3\textbf{r}_4), ~
\textbf{x}_4=\frac{1}{\sqrt{2}}(\textbf{r}_5-\textbf{r}_6) \nonumber \\
\textbf{x}_5 &= & \sqrt{\frac{4}{{3}}}(\textbf{r}_{1234} - \textbf{r}_{56}),
\label{coordinate-1}
\end{eqnarray}
with $\textbf{r}_{1234}= (\textbf{r}_1+\textbf{r}_2+\textbf{r}_3+\textbf{r}_4)/4$ and
 $\textbf{r}_{56}=(\textbf{r}_5+\textbf{r}_6)/2$. Then the  color-flavor-spin  part of the wave function should obey \{1234\}\{56\} symmetry in order to satisfy the  Pauli exclusion principle.

We introduce two additional Jacobi coordinates for the two-baryon configuration
 such as  (qqq)+(qss) and (qqs)+(qqs) to extract the relative  kinetic energy.
Specifically, for the final state involving (qqq)+(qss), we transform the coordinates in Eq.(\ref{coordinate-1}) to the following baryon-baryon coordinates:
$\textbf{y}_1=\frac{1}{\sqrt{2}}(\textbf{r}_1-\textbf{r}_2)$, $\textbf{y}_2=\frac{1}{\sqrt{6}}(\textbf{r}_1+\textbf{r}_2-2\textbf{r}_3)$,
$\textbf{y}_3=\frac{1}{\sqrt{2}}(\textbf{r}_5-\textbf{r}_6)$,
$\textbf{y}_4=\frac{1}{\sqrt{6}}(\textbf{r}_5+\textbf{r}_6-2\textbf{r}_4)$,
 $\textbf{y}_5=\sqrt{\frac{3}{2}}(\textbf{r}_{123}-\textbf{r}_{456})$, where
 $\textbf{r}_{123}= (\textbf{r}_1+\textbf{r}_2+\textbf{r}_3)/3$,
 $ \textbf{r}_{456}= ({m_u}\textbf{r}_4+{m_s}\textbf{r}_5+{m_s}\textbf{r}_6)/(m_u+2m_s)$.
Similarly, for the final state involving (qqs)+(qqs), we transform the coordinates in Eq.(\ref{coordinate-1}) to the following baryon-baryon coordinates:
$\textbf{z}_1=\frac{1}{\sqrt{2}}(\textbf{r}_1-\textbf{r}_5)$,
$\textbf{z}_2=\sqrt{\frac{2}{3}}( \textbf{r}_{15}-\textbf{r}_2)$,
$\textbf{z}_3=\frac{1}{\sqrt{2}}(\textbf{r}_3-\textbf{r}_4)$,
$\textbf{z}_4=\frac{1}{\sqrt{6}}(\textbf{r}_3+\textbf{r}_4 - 2\textbf{r}_6)$,
 $\textbf{z}_5=\sqrt{\frac{3}{2}}(\textbf{r}_{125}-\textbf{r}_{346})$ where
  $ \textbf{r}_{15}= ({m_u}\textbf{r}_1+{m_s}\textbf{r}_5)/(m_u+m_s)$,
  $ \textbf{r}_{125}= ({m_u}\textbf{r}_1+{m_u}\textbf{r}_2+{m_s}\textbf{r}_5)/(2m_u+m_s)$, and
 $ \textbf{r}_{346}= ({m_u}\textbf{r}_3+{m_u}\textbf{r}_4+{m_s}\textbf{r}_6)/(2m_u+m_s)$.
 We use  these coordinates also when we evaluate the two-body interaction between  the
   $u$ quark and the $s$ quark.
Then the relative kinetic energies in terms of these two coordinate systems for  (qqq)+(qss) or (qqs)+(qqs) are
\begin{align}
  K_{\mathrm{rel}}=\frac{1}{2\mu_1}P_{\textbf{y}_5}^2 \quad \mathrm{or} \quad K_{\mathrm{rel}}=\frac{1}{2\mu_2}P_{\textbf{z}_5}^2 \label{K_rel}
\end{align}
with $\mu_1=\frac{m_u(m_u+2m_s)}{2m_u+m_s}$ and $\mu_2=\frac{2m_u+m_s}{3}$.
 Note that $P_{\textbf{y}_5}$ and  $P_{\textbf{z}_5}$ are the conjugate momenta of $\textbf{y}_5$ and  $\textbf{z}_5$, respectively.

\section{Interaction ratios for flavor SU(3) symmetric case}
\label{sec:CSI}

Let us now consider  the flavor SU(3) symmetric case where
the matrix element shown below is  a basic quantity to determine $V_{\rm CQM}$:
\begin{align}
  {\cal X} & \equiv
  -\sum_{i< j}^N \langle \lambda_i^c \lambda_j^c\ \sigma_i \cdot \sigma_j \rangle  \nonumber \\
  &
  = N(N-10)+\frac{4}{3}S(S+1)+ 4C_{\rm f}+2C_{\rm c}.
\label{color-spin-formula}
\end{align}
Here   $C_{\rm f}$ and $C_{\rm c}$ are
the first kind of Casimir operators of flavor and color for the $N$-quark system, respectively \cite{Aerts:1977rw}.\footnote{More explicitly,
$C_{\rm f}=\frac{1}{3} (p^2+q^2+3(p+q)+pq)$ with
 $(p,q)=(0,0)$ for flavor SU(3)  singlet, (1,0) for triplet, (1,1) for octet,   (2,2) for 27-plet,   (3,0) for decuplet,
  and (0,3) for anti-decuplet. The same holds for $C_{\rm c}$ in color SU(3).  For the simplest case with $N=1$,
  we have $(p,q)=(1,0)$ in both color and flavor, so that $C_{\rm f}=C_{\rm c}=4/3$.}
 Assuming  that the size of the quark wave-function of the six-quark system
and that for each 3-quark system are the same for simplicity, we obtain, with a common
constant $\gamma$,
\begin{eqnarray}
 V_{\rm CQM}(F_\ell)  = \gamma ({\cal X}_{6q}-({\cal X}_{B}+{\cal X}_{B'})) .
  \end{eqnarray}
  In the following, we consider the ratios  between $V_{\rm CQM}(F_\ell)$  to
   get rid of the constant $\gamma$ in the flavor SU(3) symmetric limit:
   \begin{eqnarray}
 {\cal R}_{\ell}^{\rm CQM} = \frac{V_{\rm CQM} (F_\ell)}{V_{\rm CQM}(F_{27})} . \label{eq:R-CQM}
  \end{eqnarray}
  Similarly,   the  ratios for   baryon-baryon  potential at $r=0$ can be introduced as
     \begin{eqnarray}
 {\cal R}_{\ell}^{\rm LQCD} =  \frac{V_{\rm LQCD} (F_\ell)}{V_{\rm LQCD}(F_{27})} . \label{eq:R-LQCD}
  \end{eqnarray}
As mentioned in the Introduction,
 the quark wave function dependence in CQM   and the interpolating operator dependence in LQCD
 are expected to cancel independently  in  such ratios,
   so that the comparison can be made with less ambiguities.

In Table \ref{BB-interaction-1},  we compare  ${\cal R}_{\ell}^{\rm CQM}$ from the color-spin interaction in the
 favor SU(3) case  and  ${\cal R}_{\ell}^{\rm LQCD} $  obtained from the currently available lattice data
 in the flavor SU(3) case  \cite{Inoue:2011ai} with the    pseudo-scalar meson mass $M{\rm ps} \simeq 469$ MeV  and the octet baryon mass $M_{B} \simeq  1161$ MeV.   The errors in the parentheses for  ${\cal R}_{\ell}^{\rm LQCD} $  reflect the
 combined  statistical and systematic errors estimated from the LQCD data at the Euclidean time $t=11$ and $12$.
 One finds that the sign and magnitude of these ratios are qualitatively consistent with each other between CQM and LQCD.

\begin{center}
\begin{table}
  \begin{tabular}{|c|c|c|c|c|c|}
    \hline
   $ F_\ell $ & $ F_1 $ & $ F_{27} $ & $ F_{10} $ &  $ F_{\overline{10}} $ & $F_8 $ \\
        \hline \hline
   $ {\cal R}_{\ell}^{\rm CQM}  $ & $-$0.33 & 1  & 0.78 & 0.78 & 0.28 \\
    \hline
   $ {\cal R}_{\ell}^{\rm LQCD} $  & $-$0.53(3)  & 1
    & 0.93(1)   & 0.81(1)   & 0.20(1)   \\
    \hline
  \end{tabular}
  \caption{Comparison of ratios of the color-spin interactions at short distance $R_\ell$ between the constituent quark model (CQM) and the lattice QCD data (LQCD) in the flavor SU(3)  symmetric case  \cite{Inoue:2011ai}. }
  \label{BB-interaction-1}
\end{table}
\end{center}

\section{Interaction ratios for flavor SU(3) non-symmetric case}

 Let us now examine the flavor SU(3) non-symmetric case where
 the color-spin factor receives additional constituent quark mass dependence   as follows \cite{SilvestreBrac:1992yg},
\begin{align}
  {\cal Y}_\ell  \equiv
  -\sum_{i< j}^6 \frac{1}{m_i m_j} \langle \lambda_i^c \lambda_j^c \ \sigma_i \cdot \sigma_j \rangle_{{F}_\ell} .
\end{align}
Their explicit forms read
\begin{eqnarray}
& & {\cal Y}_1 =-\frac{5}{m_u^2}-\frac{22}{m_u m_s}+\frac{3}{m_s^2}, \nonumber \\
& & {\cal Y}_{27}=-\frac{13}{3m_u^2}+\frac{26}{3m_u m_s}+\frac{11}{3m_s^2}, \nonumber \\
& & {\cal Y}_{10}=\frac{98}{27m_u^2}-\frac{100}{27m_u m_s}+\frac{74}{27m_s^2}, \nonumber \\
& & {\cal Y}_{\overline{10}}=\frac{106}{27m_u^2}-\frac{116}{27m_u m_s}+\frac{82}{27m_s^2}, \nonumber \\
& & {\cal Y}_8=-\frac{16}{3m_u^2}-\frac{20}{3m_u m_s}+\frac{8}{3m_s^2}.
\label{broken-cs-factors}
\end{eqnarray}
The  numerical coefficients in Eq.~(\ref{broken-cs-factors}) are obtained by combining the
  color-spin factor  for an $(i,j)$-pair, $-\lambda^c_i \lambda^c_j \sigma_i \cdot \sigma_j$,
 with the probability of
 finding an $(i,j)$-pair in the six-quark state, $P_{ij}(F_{\ell}) $, given in Table \ref{probability-diquark}.
 For example,
 the color-spin factor $-5$ in front of  $1/m_u^2$ for  ${\cal Y}_1$   in Eq.~(\ref{broken-cs-factors}) is
 obtained by using the
four factors for $i,j$=1-4 in the 6th and 7th rows of Table \ref{probability-diquark} as follows,
\begin{align}
  \left(\frac{1}{4}\cdot (-8)+\frac{1}{4}\cdot (-\frac{4}{3})+\frac{3}{8}\cdot \frac{8}{3}+\frac{1}{8}\cdot 4\right)\cdot {4\choose 2}=-5,
\label{example}
\end{align}
where ${4\choose 2}$ is the number of light quark  pairs inside the six-quark system.
Other coefficients can be obtained similarly by further noting that the number of light-strange (strange-strange) quark pairs is 8 (1).
  The explicit form of $P_{ij}(F_{\ell}) $ is given by
\begin{align}
  P_{ij}(F_{\ell})=  \langle \Psi_{F_{\ell}}| {\cal \hat{P}}_{ij}   | \Psi_{F_{\ell}} \rangle,
\label{probability}
\end{align}
with a projection operator,
${\cal \hat{P}}_{ij} \equiv |\psi^{\rm d}_{ij}  \rangle \langle \psi^{\rm d}_{ij} | \otimes \mathbf{1}$,
 where $\psi^{\rm d}_{ij} $ is  the  wavefunction of the relevant diquark that satisfies a certain symmetry property represented in
 Table \ref{probability-diquark}, while  $\mathbf{1}$ is a unit operator acting on the particles other than $i$ and $j$.

There are a few points to be noted. First,  the isospin determines whether the color-spin factors coming from the two light quarks, shown in the first terms in Eq.~(\ref{broken-cs-factors}), are attractive ($I$=0) or repulsive ($I$=1).
As can be seen from the $P$'s appearing in the second and third columns of Table \ref{probability-diquark}, all the $I$=0
flavor states have the same contributions to the color-spin factor  from the attractive light-light diquarks, which are also larger than those from the $I$=1 case.
Furthermore, the large attraction in the $F_1$ channel as seen in the second term in the first line of Eq.~(\ref{broken-cs-factors})
follows from the large additional attraction between the light-strange diquarks, which can be seen to follow from the relatively large contribution from  the attractive maximally antisymmetric color-spin configuration  often refereed to as the ``good diquark".
Finally, as can be seen in Table \ref{probability-diquark},
the light-strange diquarks have mixed flavor symmetry  as they do not have to satisfy Pauli principle in flavor SU(3) breaking case.

\begin{center}
\begin{table}
  \begin{tabular}{|c|c|c|c|c|c|c|c|c|c|c|c|c|c|c|}
    \hline
    & \multicolumn{4}{|c|}{$i,j=1$-$4$} & \multicolumn{4}{|c|}{$i=1$-$4,\ j=5,6$} & \multicolumn{2}{|c|}{$i,j=5,6$} \\
    \hline  \hline
    $\psi^{\mathrm{d}}_{ij}$ & \multicolumn{4}{|c|}{$A$} & \multicolumn{4}{|c|}{$M$} & \multicolumn{2}{|c|}{$A$}  \\
    \hline
    color & ~ $A$~ & ~$S$~ & ~$A$~ &~ $S$ ~ &~ $A$~ & ~$S$~ & ~$A$~ & ~$S$~ & ~$A$~ & ~$S$~  \\
    \hline
    flavor & $A$ & $A$ & $S$ & $S$ & \multicolumn{4}{|c|}{$M$} & $S$ & $S$  \\
    \hline
    spin & $A$ & $S$ & $S$ & $A$ & $A$ & $S$ & $S$ & $A$ & $S$ & $A$  \\
    \hline
    -$\lambda^c_i \lambda^c_j \sigma_i \cdot \sigma_j$ & -8 & -$\frac{4}{3}$  & $\frac{8}{3}$ & 4 & -8 & -$\frac{4}{3}$  & $\frac{8}{3}$ & 4 & $\frac{8}{3}$ & 4 \\
    \hline
    $P_{ij}(F_1)$ & $\frac{1}{4}$ & $\frac{1}{4}$ & $\frac{3}{8}$ & $\frac{1}{8}$ & $\frac{3}{8}$ & $\frac{3}{8}$ & $\frac{3}{16}$& $\frac{1}{16}$& $\frac{3}{4}$ & $\frac{1}{4}$  \\
    \hline
    $P_{ij}(F_{27})$ & $\frac{1}{4}$ & $\frac{1}{4}$ & $\frac{7}{24}$ & $\frac{5}{24}$ & $\frac{1}{8}$ & $\frac{1}{8}$ & $\frac{9}{16}$& $\frac{3}{16}$& $\frac{1}{4}$ & $\frac{3}{4}$  \\
    \hline
    $P_{ij}(F_{10})$ & $\frac{5}{36}$ & $\frac{7}{36}$ & $\frac{14}{27}$ & $\frac{4}{27}$ & $\frac{5}{24}$ & $\frac{7}{24}$ & $\frac{11}{36}$& $\frac{7}{36}$& $\frac{17}{18}$ & $\frac{1}{18}$  \\
    \hline
    $P_{ij}(F_{\overline{10}})$ & $\frac{5}{36}$ & $\frac{7}{36}$ & $\frac{13}{27}$ & $\frac{5}{27}$ & $\frac{5}{24}$ & $\frac{7}{24}$ & $\frac{13}{36}$& $\frac{5}{36}$ & $\frac{3}{18}$ & $\frac{5}{18}$  \\
    \hline
    $P_{ij}(F_8)$ & $\frac{1}{4}$ & $\frac{1}{4}$ & $\frac{5}{12}$ & $\frac{1}{12}$ & $\frac{7}{32}$ & $\frac{11}{32}$ & $\frac{9}{32}$& $\frac{5}{32}$ & 1 &0  \\
    \hline
  \end{tabular}
  \caption{Color-spin factor and the  probability $P_{ij}$ for $(i,j)$ diquark pairs with  $S$,$A$ and $M$ representing symmetric ($S$),
   antisymmetric ($AS$)  and mixed ($M$) combinations.   }
  \label{probability-diquark}
\end{table}
\end{center}

To calculate the interaction energy in each flavor channel $F_{\ell}$,
we need to subtract $E_{BB'}$  in  Eq.~(\ref{binding})  from $\langle H \rangle_{F_\ell}$
 with appropriate SU(3) CG coefficients ~\cite{deSwart:1963pdg}:
\begin{align}
 V_{\rm CQM} ({F}_1) &=\langle H \rangle_{F_1} - \left[ \frac{1}{8}E_{\Lambda \Lambda}+\frac{3}{8}E_{\Sigma \Sigma}+\frac{1}{2}E_{N\Xi} \right], \nonumber\\
 V_{\rm CQM}({F}_{27})&=\langle H \rangle_{F_{27}}  - \left[ \frac{27}{40}E_{\Lambda \Lambda}+\frac{1}{40}E_{\Sigma \Sigma}+\frac{3}{10}E_{N\Xi} \right], \nonumber\\
 V_{\rm CQM}({F}_{10})&=\langle H \rangle_{F_{10}} - \left[ \frac{1}{2}E_{\Sigma \Lambda}+\frac{1}{6}E_{\Sigma \Sigma}+\frac{1}{3}E_{N\Xi} \right], \nonumber\\
 V_{\rm CQM}({F}_{\overline{10}})&=\langle H \rangle_{F_{\overline{10}}}  - \left[ \frac{1}{2}E_{\Sigma \Lambda}+\frac{1}{6}E_{\Sigma \Sigma}+\frac{1}{3}E_{N\Xi} \right], \nonumber\\
 V_{\rm CQM}({F}_8)&=\langle H \rangle_{F_8}  - E_{N\Xi}.
 \label{eq:V_CQM}
\end{align}

 The total energy of the six quarks $\langle H \rangle_{F_{\ell}}$ is given by calculating each term in
Eq.(\ref{eq:H6q}) and  minimizing the sum with respect to  the variational parameters $a_1,a_4,a_5$ in Eq.(\ref{eq:a145}).
 Resulting  total energy and  its decomposition  are shown in Table~\ref{dibaryon-mass} for each flavor state.

\begin{center}
\begin{table}
  \begin{tabular}{|c|c|c|c|c|c|c|c|}
    \hline
     $F_\ell$ & $\langle H \rangle_{6q}$  & $K$ & $E_C$ & $E_{CS}$ & $a_1$ & $a_4$ & $a_5$  \\
    \hline \hline
     $F_1$ & 2596.2 & 1717.9 & -1447.4 & -310.3 & 2.17 & 3.11 & 2.72 \\
    \hline
   $F_{27}$ & 2900.7 & 1373.6 & -1142.9 & 34.0 & 1.78 & 1.95 & 2.40 \\
    \hline
    $F_{10}$ & 2889.2 & 1372.2 & -1154.7 & 35.7 & 1.74 & 2.86 & 1.85 \\
    \hline
    $F_{\overline{10}}$ & 2900.3 & 1377.4 & -1145.2 & 32.2 & 1.73 & 2.60 & 2.14 \\
    \hline
    $F_8$ & 2735.2 & 1526.0 & -1306.7 & -120.2 & 2.0 & 3.01 & 1.91 \\
    \hline
  \end{tabular}
  \caption{Energies of the six-quark states in MeV unit. $K$, $E_C$ and $E_{CS}$ represent the total kinetic energy, confinement potential energy and color-spin potential energy, respectively. $a_1$, $a_4$ and $a_5$ are variational parameters in $\mathrm{fm}^{-2}$ unit. }
  \label{dibaryon-mass}
\end{table}
\end{center}

Corresponding LQCD data  in each flavor channel $F_{\ell}$
 can be obtained by relating the results of the baryon mass eigenstates to that of six-quark flavor eigenstates \cite{Sasaki:2015ifa}
   with the same SU(3)  CG coefficients as those  in  Eq.(\ref{eq:V_CQM}).
     By using  LQCD data for two baryons  at nearly physical quark masses corresponding to
 $(M_{\pi}, M_K, M_N, M_{\Lambda}, M_{\Sigma}, M_{\Xi}) \simeq (146, 525, 956, 1121, 1201, 1328)\ {\rm MeV}$,
 the diagonal parts of the potentials in the flavor-space $V_{\rm LQCD} (F_\ell)$  have been obtained in \cite{Inoue:2016qxt}.

 In Table \ref{BB-interaction-2},  we compare the ratios  ${\cal R}_{\ell}^{\rm LQCD}$ obtained in this way and
  ${\cal R}_{\ell}^{\rm CQM}$ obtained from Eq.(\ref{eq:R-CQM}) using Eq.(\ref{eq:V_CQM}).
  The errors in the parentheses for  ${\cal R}_{\ell}^{\rm LQCD} $  reflect the
 combined  statistical and systematic errors estimated from the LQCD data at the Euclidean time $t=11$ and $12$.
  Again, we find that the sign and magnitude of these ratios are qualitatively  consistent between CQM and LQCD.

\begin{center}
\begin{table}[h]
  \begin{tabular}{|c|c|c|c|c|c|}
    \hline
   $ F_\ell $ & $ F_1 $ & $ F_{27} $ & $ F_{10} $ &  $ F_{\overline{10}} $ & $F_8 $ \\
    \hline \hline
      $ {\cal R}_{\ell}^{\rm CQM} $  & $-$0.19 & 1 & 0.77 & 0.80 & 0.42 \\
    \hline
    ${\cal R}_{\ell}^{\rm LQCD} $ & $-$0.43(14)   & 1   & 0.97(4)   & 0.82(1)   & 0.26(2)   \\
    \hline
  \end{tabular}
\caption{Comparison of the ratios of short distance interactions $R_\ell$ between the constituent quark model (CQM) and the lattice QCD data (LQCD) in the flavor SU(3)  non-symmetric case  \cite{Inoue:2016qxt}.} \label{BB-interaction-2}
\end{table}
\end{center}

\begin{figure}[htbp]
\centering
         \includegraphics[scale=0.9]{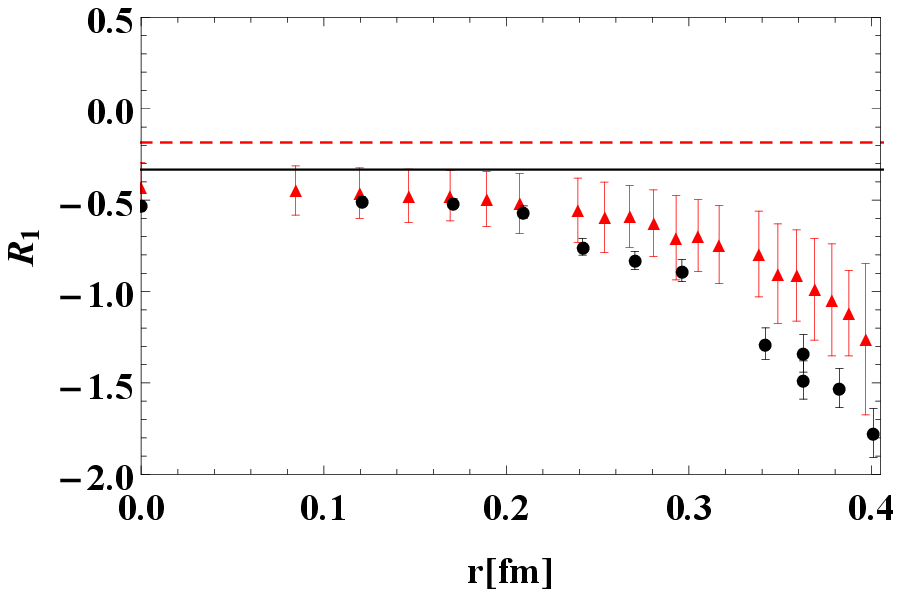}
\caption[]{The Interaction ratios ${\cal R}_{1}^{\rm LQCD}$ evaluated for finite $r$
  are plotted by black dots (SU(3) symmetric case  \cite{Inoue:2011ai}) and
 red triangles (SU(3) non-symmetric case  \cite{Inoue:2016qxt}).
 The quark model results evaluated  at $r=0$  are shown by the  black solid line (SU(3) symmetric case)
and the red dashed line (SU(3) non-symmetric case for comparison.}
\label{F1}
\end{figure}

\begin{figure}[htbp]
\centering
         \includegraphics[scale=0.9]{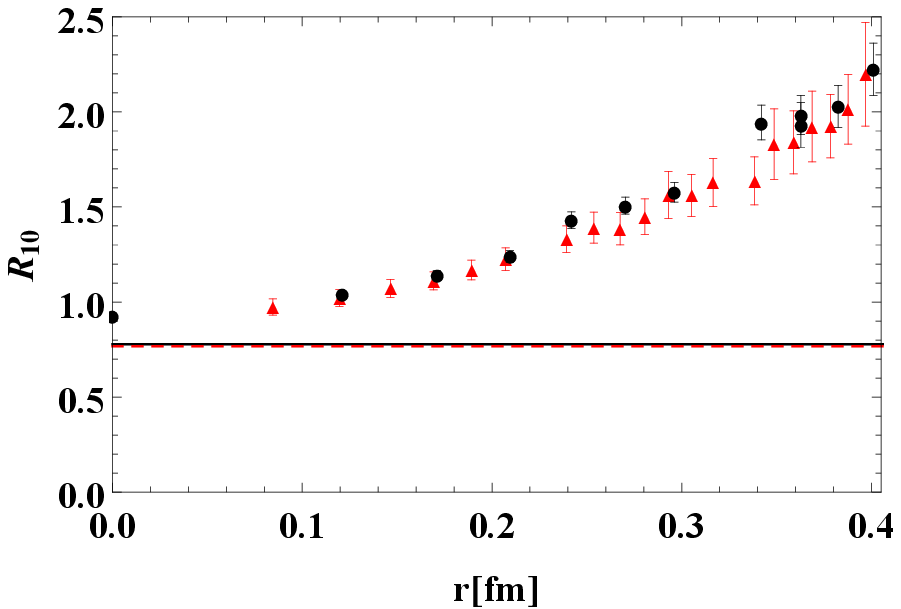}
\caption[]{Same as Fig.~\ref{F1}  for ${\cal R}_{10}^{\rm LQCD}$.}
\label{F10}
\end{figure}

\begin{figure}[htbp]
\centering
         \includegraphics[scale=0.9]{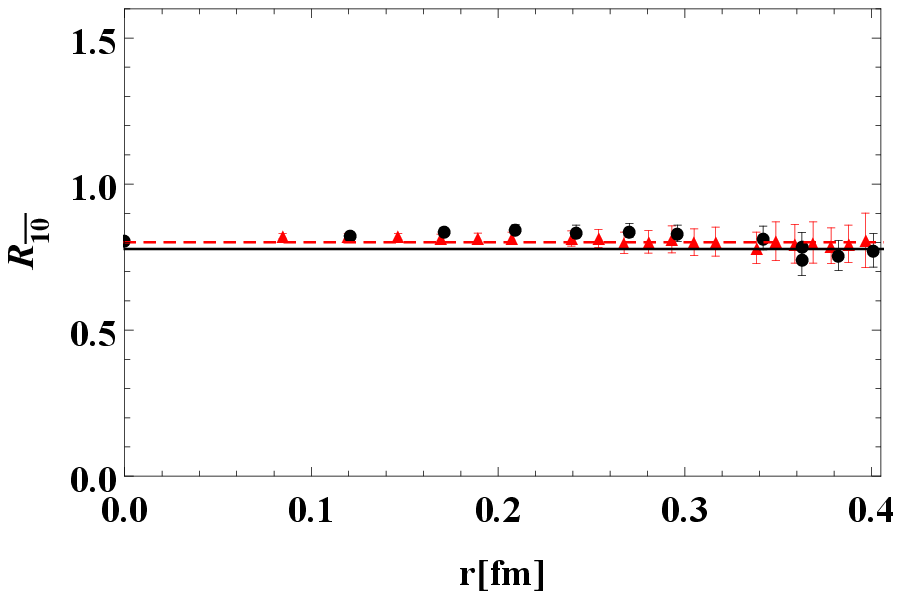}
\caption[]{Same as Fig.~\ref{F1} for ${\cal R}_{\overline{10}}^{\rm LQCD}$.}
\label{F10bar}
\end{figure}

\begin{figure}[htbp]
\centering
         \includegraphics[scale=0.9]{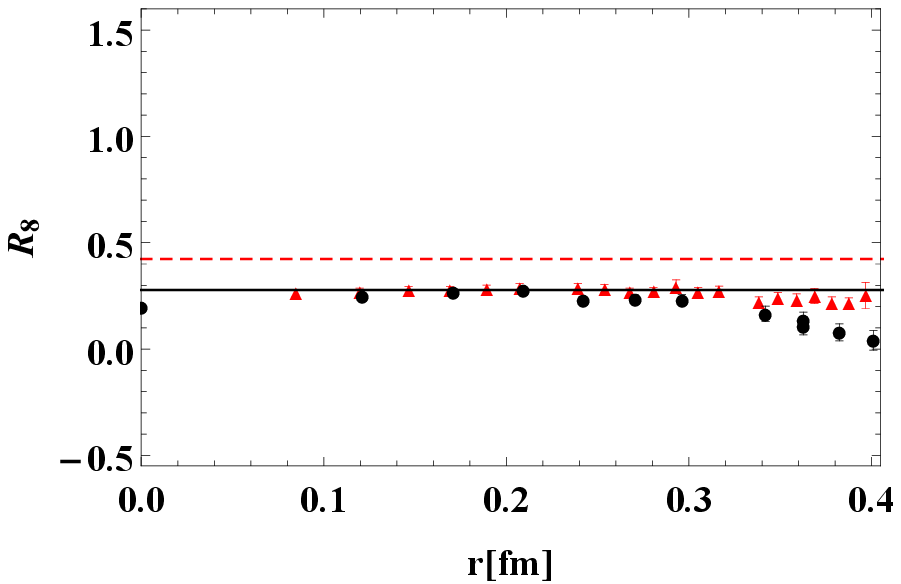}
\caption[]{Same as Fig.~\ref{F1} for ${\cal R}_{8}^{\rm LQCD}$.}
\label{F8}
\end{figure}

To see how  the interaction ratios in LQCD change as a function of $r$ at short distances,
we plot  ${\cal R}_{\ell}^{\rm LQCD}$ obtained at finite $r$ (black dots and red triangles) in Figs.~\ref{F1}-\ref{F8}
 together with ${\cal R}_{\ell}^{\rm CQM}$ (black solid lines and red dashed lines) evaluated at $r=0$.
  The error bars contain both statistical and systematic errors estimated from the LQCD data at $t=11$ and 12.
  One finds that the interaction ratio introduced in the present paper is rather insensitive to the quark masses and
   flavor SU(3) breaking at short distances, $r < 0.2$ fm.


\section{summary and concluding remarks}

Great progress has been made recently towards understanding nuclear force starting directly from QCD using lattice gauge theory.  Traditionally, nuclear force is divided into short range repulsion, intermediate attraction, and long range pion exchange.  In this work we have shown that the short range part of the baryon-baryon  interaction in different flavor channels extracted from lattice calculation  can be quantitatively understood in terms of Pauli principle and
 the quark-quark interaction.
In channels where a compact six-quark state are Pauli blocked, the baryon interaction is highly repulsive.  However, when the channels are Pauli allowed, the interaction can either be attractive or repulsive mainly depending on the sum of color-spin dynamics between quarks.
In fact, the quark dynamics
responsible for short range attraction between two baryons in certain channels provide a reason for the possible existence of
 dibaryon states (either compact or molecular type)  in the same quantum numbers, such as the elusive H-dibaryon.
Also, by analyzing the color-flavor-spin wave function and all possible diquark configuration contributing to a given six-quark state with two strange quarks, we have shown that the large attraction originates from the light-light diquark in the iso-singlet and spin-singlet channel as well as the light-strange diquark in the spin-singlet channel.  This  implies the crucial role of  "good diquarks" in the multi-quark system with strangeness.

There are still rooms for improvement in our work.  For more quantitative comparison between CQM and LQCD, the contributions from non-zero quark orbital states should be taken into account.  Appendix A contains useful matrix elements for such extended analysis. Furthermore, to go beyond the discussions limited to short distance we have to properly include further non-perturbative quark dynamics and/or the effects of  flavor-spin dynamics through pseudo-scalar meson exchange~\cite{Glozman:1995fu}.  Such issues will be discussed in the future.

\

\

\section*{Acknowledgments}
The work by SHL was supported by Samsung Science and Technology Foundation under Project Number SSTF-BA1901-04.  This work by AP was supported by the Korea National Research
Foundation under the grant number 2018R1D1A1B07043234.
 The work by T.H. was supported by JSPS Grant-in-Aid for Scientific Research (S), No. 18H05236.
 We thank H. Nemura for carefully reading the manuscript.

\appendix

\section{Beyond s-wave orbital for six-quark state in CQM}

In the main text, we focus only on the s-wave orbitals for the 6-quark systems to extract the interaction $V_{\rm CQM}$.
When the two baryons overlap with each other,  however,  there
  arise  four possible orbital states in general. They are
 characterized by the   Young tableau as,
\begin{eqnarray}
[3]_O \times [3]_O=[6]_O +[51]_O +[42]_O +[33]_O . \nonumber
\end{eqnarray}
 Flavor-spin structure associated with each orbital state for color-singlet 6-quark is shown in  the first row of TABLE \ref{mixed-orbital}.
 Then  the color-spin matrix element $\chi(F_{\ell})$ for $N=6$  (defined in the first equality in Eq.(\ref{color-spin-formula}))
  can be evaluated as  summarized in TABLE \ref{mixed-orbital}.
   The column for $[6]_O \otimes [33]_{FS}$ corresponds to the matrix elements employed  in the text.

\begin{widetext}
\begin{center}
\begin{table}[b]
\begin{tabular}{|c|c|c|c|c|c|}
  \hline
  & $[6]_O\otimes[33]_{FS}$ & $[51]_O\otimes[42]_{FS}$ & $[42]_O\otimes[51]_{FS}$ & $[42]_O\otimes[33]_{FS}$ & $[33]_O\otimes[42]_{FS}$ \\
  \hline \hline
  ${\cal X}_{6q}(F_1)$ & -24 & p.f. & p.f. & -14 & p.f. \\
  \hline
  ${\cal X}_{6q}(F_{27})$ & 8 & p.f. & $\frac{16}{9}$ & $-\frac{38}{9}$ & p.f.  \\
  \hline
  ${\cal X}_{6q}(F_{10})$ & $\frac{8}{3}$ & $\frac{16}{15}$ & $-\frac{128}{9}$ & $-\frac{26}{9}$ & $-\frac{4}{3}$ \\
  \hline
  ${\cal X}_{6q}(F_{\overline{10}})$ & $\frac{8}{3}$ & p.f.   & $\frac{16}{9}$ & $-\frac{26}{9}$ &  p.f. \\
  \hline
  ${\cal X}_{6q}(F_8)$ & $-\frac{28}{3}$ & $\frac{2}{15}(-67 \pm 3\sqrt{241})$ & $-\frac{80}{9}$ & $-\frac{59}{9}$ & $\frac{1}{6}(-41\pm3\sqrt{457})$  \\
  \hline
\end{tabular}
\caption{The expectation value of color-spin interaction with respect to the 6-quark systems with mixed orbital symmetry in flavor SU(3) symmetric case. The elements with ``p.f." in the Table correspond to the Pauli forbidden states.}
\label{mixed-orbital}
\end{table}
\end{center}
\end{widetext}

\end{document}